\documentclass{natureprintstyle}
\usepackage{ifpdf} 
\ifpdf
  \pdfoutput=1  % arxiv.org needs this in the first 5 lines to detect pdflatex   
 \usepackage[pdftex]{graphicx}         
  \usepackage{pdftricks}
  \usepackage{epstopdf}
  \usepackage[colorlinks, bookmarks, breaklinks, pdftitle={},pdfauthor={Anil C. Seth}]{hyperref}
   
  \hypersetup{linkcolor=black,citecolor=black,filecolor=black,urlcolor=blue}       
  \usepackage[kerning,final]{microtype}  % pdfsync does not work with kerning
\else                                
  \usepackage[dvips]{graphicx} 
\fi

\usepackage{amssymb}
\usepackage{relsize}

\def\aj{{Astron.~J.}}                   % Astronomical Journal
\newcommand{\araa}{Annual Rev. Astron. Astrophys.}
\def\apj{{Astrophys.~J.}}                 % Astrophysical Journal
\def\apjl{{Astrophys.~J.~Let.}}                % Astrophysical Journal, Letters
\def\apjs{{Astrophys.~J.~Supp.}}               % Astrophysical Journal, Supplement
\def\aap{{Astron. Astrophys.}}                % Astronomy and Astrophysics
\def\mnras{{Mon. Not. R. Astron. Soc.}}             % Monthly Notices of the RAS
\def\pasp{{PASP}}               % Publications of the ASP
\def\nat{{Nature}}              % Nature
\def\farcs{\hbox{$.\!\!^{\prime\prime}$}}

\def\arcsec{\hbox{$^{\prime\prime}$}}

\title{A Supermassive Black Hole in an Ultracompact Dwarf Galaxy}

\begin{document}

\author{Anil C.~Seth$^1$, Remco van den Bosch$^2$, Steffen Mieske $^3$, Holger Baumgardt$^4$, Mark den Brok$^1$, Jay Strader$^5$, Nadine Neumayer$^{2,6}$, Igor Chilingarian$^{7,8}$, Michael Hilker$^6$, Richard McDermid$^{9,10}$, Lee Spitler$^{9,10}$, Jean Brodie$^{11}$,  Matthias J.~Frank$^{12}$, Jonelle L.~Walsh$^{13}$}

\maketitle

\begin{affiliations}
 \item Department of Physics \& Astronomy, University of Utah, 115 South 1400 East, Salt Lake City, UT 84112, USA
 \item Max-Planck Institut f\"ur Astronomie, K\"onigstuhl 17, D-69117 Heidelberg, Germany
 \item European Southern Observatory, Alonso de Cordova 3107, Vitacura, Santiago, Chile
 \item School of Mathematics and Physics, University of Queensland, St Lucia, QLD 4072, Australia
 \item Department of Physics \& Astronomy, Michigan State University, East Lansing, MI 48824, USA
 \item European Southern Observatory, Karl-Schwarzschild-Str. 2, 85748 Garching bei M\"unchen, Germany
 \item Smithsonian Astrophysical Observatory, 60 Garden St. MS09, Cambridge MA 02138 USA
 \item Sternberg Astronomical Institute, Moscow State University, 13 Universitetski prospect, Moscow 119992 Russia
 \item Australian Astronomical Observatory, 105 Delhi Rd, Sydney, NSW 2113, Australia 
 \item Department of Physics \& Astronomy, Macquarie University, Sydney, NSW 2109, Australia
 \item University of California Observatories and Department of Astronomy and Astrophysics, University of California, Santa Cruz, CA 95064, USA 
 \item Landessternwarte, Zentrum f\"ur Astronomie der Universit\"at Heidelberg, K\"onigsstuhl 12, D-69117 Heidelberg, Germany
 \item Department of Astronomy, The University of Texas at Austin, 1 University Station C1400, Austin, Texas 78712, USA
\end{affiliations}

\begin{abstract}
{\em Author's version of paper submitted to Nature June 21st, accepted July 29th, to appear in Sept.~18th issue at {\tt http://dx.doi.org/10.1038/nature13762}}\\

Ultracompact dwarf galaxies (UCDs) are among the densest stellar systems in the universe.  These systems have masses up to 200 million solar masses, but half light radii of just 3 $-$ 50 ~parsecs\cite{brodie11}.  Dynamical mass estimates show that many UCDs are more massive than expected from their luminosity\cite{mieske13}.  It remains unclear whether these high dynamical mass estimates are due to the presence of supermassive black holes or result from a non-standard stellar initial mass function that causes the average stellar mass to be higher than expected\cite{frank11,dabringhausen12}. Here we present the detection of a supermassive black hole in a massive UCD.  Adaptive optics kinematic data of M60-UCD1 show a central velocity dispersion peak above 100 km/s and modest rotation.  Dynamical modeling of these data reveals the presence of a supermassive black hole with mass of 21 million solar masses.  This is 15\% of the object's total mass.  The high black hole mass and mass fraction suggest that M60-UCD1 is the stripped nucleus of a galaxy. Our analysis also shows that M60-UCD1's stellar mass is consistent with its luminosity, implying many other UCDs may also host supermassive black holes.  This suggests a substantial population of previously unnoticed supermassive black holes.

\end{abstract}

%\item Description of data 
The object M60-UCD1 is the brightest ultracompact dwarf galaxy (UCD) currently known\cite{strader13} with a luminosity $L_V = 4.1 \times 10^7$~L$_\odot$ and radius $r_e = 24$~pc.  It lies at a projected distance of 6.6 kpc from the center of the massive elliptical galaxy M60 (Fig.~1), and 16.5~Mpc from us\cite{blakeslee09}.  We obtained integral field spectroscopic data between 2 and 2.4 $\mu$m of M60-UCD1 with Gemini/NIFS.  The high spatial resolution data obtained using laser guide star adaptive optics provides a clear detection of the supermassive black hole (BH).  
Modeling of the deep CO absorption bandheads at 2.3 $\mu$m enables us to measure the motions of stars at many different points across M60-UCD1.  These kinematic measurements are shown in Fig.~2.  
 Two features are particularly notable: (1) the dispersion is strongly peaked, with the central dispersion rising above 100~km/s and dropping outwards to $\sim50$~km/s, (2) rotation is clearly seen, with a peak amplitude of 40~km/s.  
%Is flattening consistent with rotation?

%\item Modeling Results
The stellar kinematics can be used to constrain the distribution of mass within M60-UCD1.  This includes being able to test whether the mass traces light, or if a supermassive BH is required to explain the central velocity dispersion peak.  We combined the stellar kinematics and imaging from the Hubble Space Telescope with self-consistent Schwarzschild models\cite{schwarzschild79,vandenbosch08, vandenbosch10} to constrain the BH mass and mass-to-light ratio ($M/L$, in solar units), shown in Figure~3. We measure a black-hole mass of 2.1$^{+1.4}_{-0.7}\times 10^7$~M$_\odot$ and $g$ band $M/L = 3.6 \pm 1$ with errors giving $1\sigma$ confidence intervals (2$\sigma$ and 3$\sigma$ contours are shown in Fig.~3).  The total stellar mass is $1.2\pm0.4\times10^{8}$~M$_\odot$.  The best fit constant $M/L$ model with no BH is ruled out with $>$99.99\% confidence.  

M60-UCD1 is the lowest mass system known to host a supermassive BH ($>10^6$~M$_\odot$) including systems with dynamical BH estimates or with broad line AGN\cite{kormendy13,reines13}. There have been tentative detections of $\sim$10$^4$~M$_\odot$ BHs in lower mass clusters\cite{gebhardt05,jalali12}.   These detections remain controversial\cite{vandermarel10,miller-jones12} and the intermediate mass BHs, if present, form a much smaller fraction of the total cluster mass than found in M60-UCD1.  
Of the 75 galaxies with reliable dynamical BH mass measurements, only one other galaxy has a BH mass fraction as high as M60-UCD1\cite{vandenbosch12,kormendy13}. A luminous and variable X-ray source was previously detected in M60-UCD1 with a maximum luminosity of $L_X = 1.3 \times 10^{38}$~ergs/s\cite{strader13}. This luminosity suggests the BH is accreting material at a rate typical of BHs in larger, more massive early type galaxies in Virgo, as well as other nearby galaxies with absorption-line dominated optical spectra \cite{gallo10,ho08}.

UCDs are thought to be either the most massive globular star clusters\cite{mieske12} or nuclei of larger galaxies that have been tidally stripped\cite{drinkwater03,pfeffer13}.  The supermassive BH that we have found at the center of M60-UCD1 provides strong evidence that it is a stripped nucleus of a once larger galaxy.  While it is possible dense star clusters can form BHs, these are expected to contain only a small fraction of the cluster's mass\cite{portegieszwart04}.  Star clusters at the center of galaxy nuclei on the other hand are known to host black holes with very high mass fractions \cite{graham09}.
Thus M60-UCD1 is the first individual UCD with explicit evidence for being a tidally stripped nucleus.  

We can estimate the properties of M60-UCD1's progenitor galaxy assuming that they follow scaling relations of present-day unstripped galaxies.  Using the known scaling between BH mass and bulge mass\cite{kormendy13}, we find a host bulge mass of 7$^{+4}_{-3} \times 10^9$~M$_\odot$.  Bulge masses are also known to correlate with the masses of their nuclear star clusters\cite{ferrarese06}.  In M60-UCD1, the surface brightness profile has two clear components\cite{strader13}, and we identify the central component as the progenitor nuclear star cluster\cite{pfeffer13} with mass $6.1\pm1.6 \times 10^7$~M$_\odot$.  This translates to a predicted bulge mass of $1.8\pm0.4 \times 10^{10}$~M$_\odot$.   Thus two independent scaling relations suggest the progenitor bulge mass is $\sim$10$^{10}$~M$_\odot$.  Given M60-UCD1's cluster environment, its progenitor was likely a lower mass elliptical galaxy that was then stripped by the massive elliptical galaxy M60 which lies at a current projected distance of just 6.6~kpc.  We have run simulations that show that it is feasible to produce M60-UCD1 by stripping a $\sim10^{10}$~M$_\odot$ elliptical galaxy progenitor on a fairly radial orbit (see supplementary information \& Extended Data Fig.~6).  We note that current $\sim$10$^{10}$~M$_\odot$ elliptical galaxies have nuclear star cluster sizes, luminosities and colors consistent with the inner component of M60-UCD1\cite{cote06,turner12}.

The detection of a supermassive BH in M60-UCD1 may be just the tip of the iceberg of the UCD BH population. Measurements of the integrated velocity dispersion in almost all UCDs with masses above $10^7$~M$_\odot$ yield dynamical mass estimates that are too high to be accounted for by a normal stellar population without a massive central
BH\cite{mieske13}.  Unlike these previous dynamical mass estimates, our dynamical modeling can separate out the gravitational influence of the BH from the contribution of the stars.  The models show that M60-UCD1's stellar populations appear normal.  In M60-UCD1 we measured a stellar dynamical mass-to-light ratio of $M/L_g = 3.6 \pm 1.0$ (1$\sigma$ errors).  This mass-to-light ratio is consistent with the stellar populations seen in lower mass globular clusters\cite{strader11} and models with a normal (Milky Way) initial mass function\cite{bastian10}.  It is also lower than the integrated dynamical mass-to-light ratio estimates in 18 of 19 UCDs above $10^7$~M$_\odot$\cite{mieske13}.  The low stellar mass-to-light ratio in M60-UCD1 is inconsistent with proposed scenarios in which a density-dependent initial mass function yields higher than normal mass-to-light ratios\cite{dabringhausen12}.  Without such a mechanism, the simplest explanation for the high dynamical mass estimates in massive UCDs is that most host supermassive BHs just like M60-UCD1.  

There is also more limited evidence for enhanced dynamical $M/L$s in lower mass UCDs.  About half of UCDs with masses between $3 \times 10^6$~M$_\odot$ and $10^7$~M$_\odot$ have higher inferred mass-to-light ratios than we see in M60-UCD1\cite{mieske13}. Taken in combination with tentative BH detections of $\sim10^4$~M$_\odot$ in Local Group globular clusters\cite{gebhardt05,jalali12} (which do not have increased $M/L$ values due to much lower BH mass fractions), these observations suggest that some lower mass UCDs may also host relatively massive BHs.

Finally, we estimate what the total population of UCD BHs might be in the local universe.  The most complete sample of known UCDs is in the Fornax cluster.  Comparing the UCD population to the population of galaxies in Fornax likely to host massive BHs, we find that UCDs may more than double the number of BHs (see supplementary information).  Thus UCD BHs could represent a large increase in the massive BH number density in the Local Universe.  Future work can test this hypothesis.  We have ongoing observing programs to obtain similar observations to the ones presented here in four additional massive UCDs, and in the most massive star clusters in the Local Group.  However, dynamical detection of black holes will be challenging in all but the brightest and nearest of objects, thus accretion signatures\cite{gultekin14} or tidal disruption events\cite{miller14b} may represent the best possibility for detecting BHs in less massive UCDs.

\begin{figure}
 \centering
 \includegraphics[width=8.9cm]{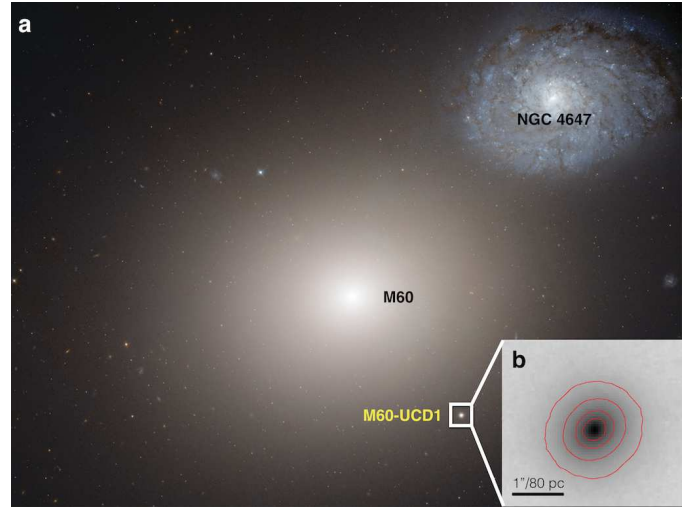}
\caption{Hubble Space Telescope image of the M60-NGC4647 system (a). M60-UCD1 is the nearly point-like image in the bottom right.  The discovery of a black hole in M60-UCD1 provides evidence that it is the tidally stripped nucleus of a once larger galaxy.  We note that NGC~4647 is at approximately the same distance as M60 but the two galaxies are not yet strongly interacting.  The inset shows a zoomed version of the $g$~band image of M60-UCD1 with contours showing the surface brightness in 1 mag/arcsec$^2$ intervals.  {\em Image Credit: NASA / ESA}} 
 \label{fig:pretty}
\end{figure}

\begin{figure}
 \centering
 \includegraphics[width=8.9cm]{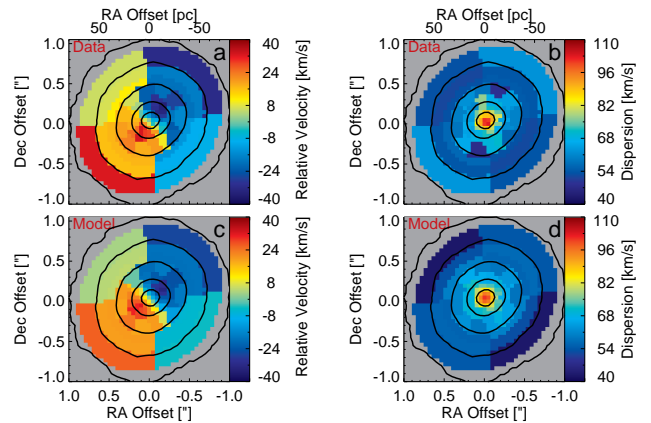}
\caption{Stellar Kinematic Maps of M60-UCD1 showing clear rotation and a dispersion peak.  Panels {\bf a} and {\bf b} show the measured radial velocities (bulk motions towards \& away from us) and velocity dispersions (random motions) of the stars in M60-UCD1 with typical errors of 6~km/s.  Black contours show isophotes in the K band stellar continuum. Kinematics are determined in each individual pixel near the center, but at larger radii the data were binned to increase signal-to-noise and enable kinematic measurements.  Panels {\bf b} and {\bf c} show the best fit dynamical model; a black hole is required to replicate the central dispersion peak.}
 \label{fig:kinematics}
\end{figure}

\begin{figure}
 \centering
 \includegraphics[width=8.9cm]{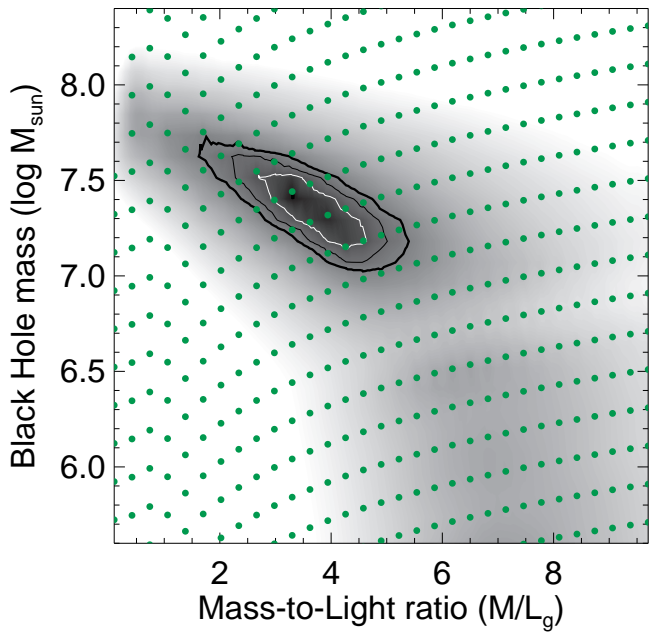}
\caption{Dynamical modelling results show the presence of a supermassive black hole.  The figure shows goodness-of-fit contours for the dynamical models of M60-UCD1 with two parameters, $g$-band mass-to-light ratio and black hole mass. The contours indicate 1$\sigma$ (white), 2$\sigma$, 3$\sigma$ (black, thick) confidence levels for two degrees of freedom. Green dots indicate discrete values of mass-to-light ratio and black hole mass at which models were fit to data.}
 \label{fig:modelchi}
\end{figure}

\begin{addendum}
 \item Based on observations obtained at the Gemini Observatory, which is operated by the Association of Universities for Research in Astronomy, Inc., under a cooperative agreement with the NSF on behalf of the Gemini partnership: the National Science Foundation (United States), the National Research Council (Canada), CONICYT (Chile), the Australian Research Council (Australia), Minist\'{e}rio da Ci\^{e}ncia, Tecnologia e Inova\c{c}\~{a}o (Brazil) and Ministerio de Ciencia, Tecnolog\'{i}a e Innovaci\'{o}n Productiva (Argentina).  Work on this paper by A.C.S.~is supported by NSF CAREER grant AST-1350389.  J.L.W.~is supported by an NSF Astronomy and Astrophysics Postdoctoral Fellowship under Award No. 1102845.  J.B.~is supported by NSF grant AST-1109878.  M.J.F.~is supported by German Research Foundation grant Ko 4161/1.  I.C.~acknowledges support by the Russian Science Foundation grant 14-22-00041."
 \item [Reprints] Reprints and permissions information is available at www.nature.com/reprints
 \item[Competing Interests] The authors declare that they have no
competing financial interests.
 \item[Correspondence] Correspondence and requests for materials
should be addressed to A.C.S.~(email: aseth@astro.utah.edu).
\item[Author Contributions] All authors helped with interpretation of the data and provided comments on the manuscript.  A.C.S.~planned observations, reduced and analyzed the data and was the primary author of the text, R.v.d.B.~created dynamical models and contributed text, S.M.~contributed text, H.B.~ran tidal stripping simulations, M.d.B.~created dynamical models and analyzed model results, J.S., N.N., and R.M.~helped plan the observations, I.C.~helped verify kinematic measurements, M.H.~and~L.S.~helped with compilation of UCD numbers.  
\end{addendum}

\renewcommand{\figurename}{Extended Data Figure}
\renewcommand{\tablename}{{\bf Extended Data Table}}
\setcounter{figure}{0}

\clearpage
%\vspace{0.5in}

\noindent {\Large {\em Methods}}

In this methods section, we discuss the details of our data (section~1), kinematics (section~2) and modeling of the light profile (section~3).  We then provide details on the dynamical modeling and discuss alternatives to a supermassive black hole in section~4.  The supplementary information (below) has additional information on our calculation of the number of BHs in UCDs compared to galaxies and details on our simulations showing that M60-UCD1 is consistent with being a $10^{10}$~M$_\odot$ galaxy tidally stripped by M60.

\section{Gemini/NIFS Data and Point Spread Function}

\hspace{1in}

The kinematic data presented here are derived from integral field spectroscopic observations of M60-UCD1 taken on February 20th, May 18th, and May 19th, 2014 using Gemini/NIFS\cite{mcgregor03} using the {\em Altair} laser guide star adaptive optics with an open loop focus model.  Gemini/NIFS provides infrared spectroscopy in $0\farcs1 \times 0\farcs04$ pixels over a 3\arcsec~field of view; our observations were taken in the $K$ band at wavelengths from 2.0 to 2.4~$\mu$m.  

The final data cube was made from a total of nine 900s on-source exposures with good image quality (four taken on Feb 20th, four on May 18th and one on May 19th).  Data were taken in an object-sky-object order.  The sky frames were taken with small $\sim$10\arcsec~offsets from the source at similar galactocentric radii within M60.  Large diagonal $\sim$1\arcsec dithers made between the two neighboring object exposures ensure that the same sky pixels were not used even when the data are binned, thus improving our signal-to-noise $S/N$.  

The Gemini/NIFS data were reduced similar to our previous work with NIFS\cite{seth10}.  Each individual data cube was corrected using an A0V telluric star (HIP58616) at similar airmass.  However, due to an error, no telluric star was observed on Feb.~20th so the telluric from May 18th was used to correct that exposure as well; we test any effects this may have on our kinematic data in the next subsection and show that they are minimal.  The Gemini NIFS pipeline was modified to enable proper error propagation and IDL codes were used to combine the final data cube including an improved outlier rejection algorithm that uses neighboring pixels to help determine bad pixels.  Each dithered data cube was shifted and combined to yield a final data cube with 0.05''$\times$0.05'' spatial pixels; a velocity offset to compensate for the differing barycentric corrections was applied to the February cubes.  The final $S/N$ in the central pixels is $\sim$60 per resolution element at $\lambda=2350$~nm.  

The instrumental dispersion of NIFS varies by $\sim$20\% across the field of view.  To determine the instrumental dispersion of each spatial pixel, the sky frame exposures were dithered and combined identically to the science images.  Using this sky data cube, we fit isolated OH sky lines in each spatial pixel using double Gaussian fits to derive the instrumental dispersion; the median FWHM was 0.421~nm.  

\renewcommand*{\thefootnote}{\fnsymbol{footnote}}

The Point Spread Function (PSF) was derived by convolving an HST ACS/WFC (PID: 12369) image to match the continuum emission in the NIFS data cube.  We used a Lucy-Richardson deconvolved version of the HST F475W image; the available F850LP image is closer in wavelength, but has a significantly more complicated and less well modeled PSF. The deconvolved F475W image was then fit to our NIFS image using the MPFIT2DFUN code\footnote{{\tt http://www.physics.wisc.edu/~craigm/idl/fitting.html}}.  A double Gaussian model was required to obtain a good fit to the PSF; the inner component has a FWHM of 0.155'' and contains 55\% of the light while the outer component has a FWHM of 0.62'' and contains 45\% of the light.  The residuals to the fit have a standard deviation of just 6\% out to a radii of 1''.  This PSF was assumed for the kinematics in all dynamical models.  

\section{Deriving Kinematics}

\hspace{1in}

The kinematics were determined by fitting the CO bandhead region (2.295-2.395 $\mu$m) to stellar templates\cite{wallace96} using the penalized pixel fitting algorithm PPXF\cite{cappellari04}.  We fit the radial velocity ($V$), dispersion ($\sigma$), skewness ($h_3$), and kurtosis ($h_4$) to the data.  Before fitting, the data were binned together using Voronoi binning\cite{cappellari03} to achieve $S/N > 35$ per resolution element in each bin.  The Voronoi bins at large radii were predominantly radial in shape, and thus beyond 0$\farcs$5 we binned spectra using elliptical sections with an axial ratio of 0.85 based on the observed ellipticity at these radii in the HST images.  
These outer bins have $S/N$ between 24 and 42 per resolution element. An example of kinematic fits in a high dispersion central pixel and low dispersion outer bin are shown in Extended Data (ED) Fig.~1.  To determine errors on the derived kinematics in each bin, Monte Carlo estimates were performed by adding Gaussian random noise to each spectral pixel in each bin, refitting the kinematics and then taking the standard deviation of the resulting data.  The central velocity of M60-UCD1 is found to be $1294\pm5$~km/s, while the integrated dispersion at $r < 0\farcs75$ is found to be $69\pm1$~km/s.  Both values are consistent with the integrated optical spectroscopy measurements\cite{strader13}.
The kinematic maps in all four velocity moments are shown in ED Fig.~2.

\begin{figure}
%\epsscale{0.7}
\begin{center}
\includegraphics[width=8.9cm]{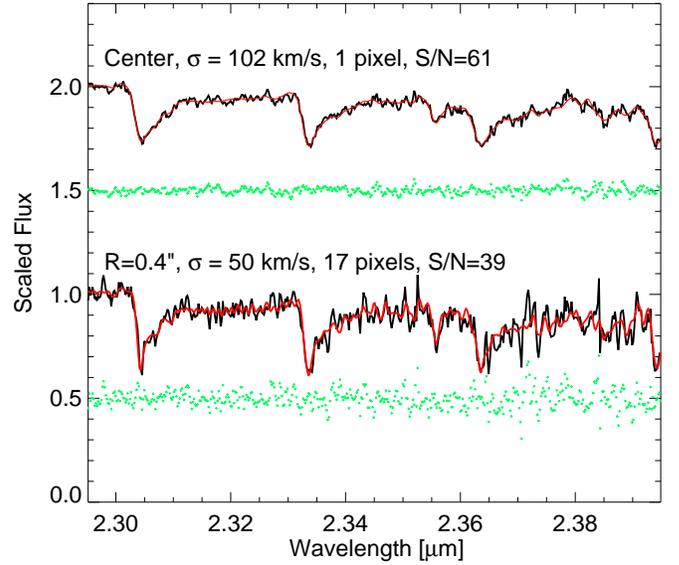}
\end{center}
\caption{Two example spectra (black lines) and their kinematic fits (red lines).  Residuals for both spectra are shown in green.  The top spectrum is from one of the central-most pixels and is the spectrum from a single $0\farcs05\times0\farcs05$ pixel.  The bottom spectrum is at a radius of $0\farcs4$ and is the sum of 17 spatial pixels. $S/N$ values are given per resolution element. The contrast in dispersion is seen very clearly, with broad smooth lines in the top spectrum and sharper lines in the bottom.  Both spectra were normalized to one; the central spectrum was then offset by +1 for visibility.  The residuals were offset by 0.5 and 1.5.}
\label{fig:specfit}
\end{figure}
The robustness of the kinematic measurements and their errors was tested by comparing the data taken on Feb.~20th with the data taken in May.  The four February and five May cubes were combined into separate final cubes.  Spectra were then extracted in the same bins as used for the full data set.  We compared the velocity and dispersion differences between the cubes to the differences expected from the errors and found that these were consistent.  More explicitly, we found that the distribution of $(V_{May}-V_{Feb})/\sqrt{Error_{May}^2 + Error_{Feb}^2}$  had a standard deviation between 0.9 and 1.1 for both velocity and dispersion measurements; the bias between the two measurements was less than the typical errors on the measurements.  We also tested for template mismatch using PHOENIX model spectra\cite{husser13} and find consistent kinematic results within the errors.  Thus we conclude that (1) our kinematic measurements are robust and (2) our kinematics errors are correctly estimated.

\begin{figure}
%\epsscale{0.7}
%\plotone{m60-ucd1_allkin.ps}
\begin{center}
\includegraphics[width=8.9cm]{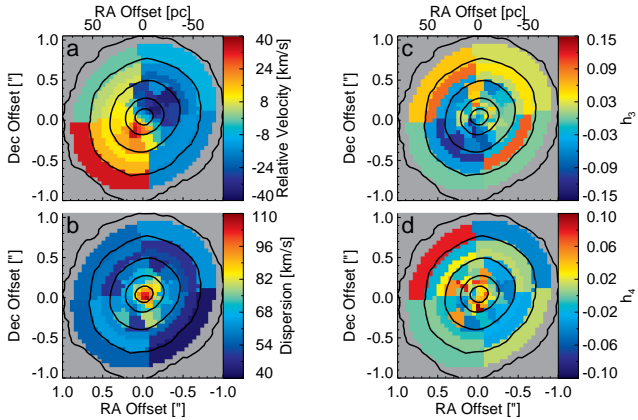}
\end{center}
\caption{The full results of kinematic fits to M60-UCD1, these include determinations of radial velocity (top-left), dispersion (bottom-left), the skewness $h_3$ (top-right), and the kurtosis $h_4$ (bottom-right).  Black contours show the $K$ band continuum at 1 mag/arcsec$^2$ intervals. The median 1$\sigma$ errors are 5.8~km/s for the velocities, 6.8~km/s for dispersion, 0.06 for $h_3$, and 0.07 for $h_4$. The skewness clearly shows the commonly seen anti-correlation with the velocity\cite{seth10b}.}
\label{fig:allkin}
\end{figure}

\section{Multi-Gaussian Expansion model of M60-UCD1}

\hspace{1in}

Archival Hubble Space Telescope data in the F475W ($g$) filter (Program ID: 12369) provides the cleanest measurement of the light distribution of M60-UCD1.  We first fitted the data to a PSF-convolved two component S\'ersic profile using methods and profiles similar to previous fits of M60-UCD1\cite{strader13}.  To enable the fitting of axisymmetric models (see \S4.1), we forced the outer nearly circular component (with axial ratio $b/a \sim$ 0.95) to be exactly circular to ensure no isophotal twist in the model due to a misalignment of the inner and outer component.  Enforcing circularity in the outer component had a negligible effect on the quality of fit compared to the previous best fit model\cite{strader13}. Our best fit axisymmetric model has:\\
Outer S\'ersic Component (circular): surface brightness $\mu_e=20.09$, effective radius $r_e=0\farcs600$, S\'ersic $n=1.20$, integrated magnitude $g=18.43$\\
Inner S\'ersic Component: Position angle $PA=-49.45$, $b/a=0.749$, $\mu_e=17.322$, $r_e=0\farcs175$, $n=3.31$, $g=18.14$\\
We generated a multi-Gaussian expansion (MGE)\cite{emsellem94,cappellari02} of this profile for use in our dynamical models.  The MGE values are shown in ED~Table~1.

We note that data in the F850LP filter of M60-UCD1 is also available, but due to the lack of a red cutoff on the filter the PSF is asymmetric, temperature dependent, and difficult to characterize. The one downside of using the F475W filter is that it is at a significantly different wavelength than the kinematic measurements.  However, there is no evidence for any color variation within the object; the inner and outer morphological components have consistent colors within 0.01 mags ($F475W-F850LP = 1.56$ \& 1.57 for the inner and outer components respectively).  The radial profile of deconvolved F475W and F850LP images is flat; any color differences are $\lesssim$0.03 mags. Thus there is no evidence for any stellar population differences within the object.\\

\section{Dynamical Modeling}

\hspace{1in}

\subsection{Modeling Details}

The most common method for measuring dynamical black hole masses is Schwarzschild orbit-superposition modeling\cite{schwarzschild79} of the stellar kinematics. Here we use a triaxial Schwarzschild code described in detail in a methods paper\cite{vandenbosch08}. To briefly summarize the method, the dynamical models are made in three steps. First, a three-dimensional luminous mass model of the stars is made by de-projecting the two-dimensional light model from the \emph{HST} image.  This is done with the MGE from the previous section, which is deprojected to construct a 3D mass distribution for the stars, assuming a constant mass-to-light ratio and a viewing angle. Second, the gravitational potential is inferred from the combination of the stellar mass and black hole mass. In a triaxial potential, the orbits conserve three integral of motions that can be sampled by launching orbits orthogonally from the x-z plane. A full set of representative orbits are integrated numerically, while keeping track of the paths and orbital velocities of each orbit. The orbit library we used for M60-UCD1 consists of 7776 orbits.  Third, we model the galaxy by assigning each orbit an amount of light, simultaneously fitting both the total light distribution and the NIFS stellar kinematics (ED~Fig.~2) including the effects of the PSF given above. Each of these steps are then repeated with different viewing angles and potentials to find the best-fitting mass distribution and confidence intervals.  The recovery of the internal dynamical structure (distribution function), intrinsic shape, and black hole mass using this code are validated in a series of papers\cite{vandeven08, vandenbosch09, vandenbosch10}. The orbit-based models are fully self-consistent and allow for all physically possible anisotropies; the models make no \emph{a priori} assumptions about the orbital configuration. 

For modeling M60-UCD1, we adopted a (nearly) oblate geometry with an intermediate axis ratio of $b/a=0.99$. In total there are three free parameters: the stellar $M/L$, the black hole mass, and the viewing angle. A total of 62 different stellar $M/L$s, sampled in linear steps, and 22 black hole masses, sampled in logarithmic steps, were modeled at 4 inclinations between 41 and 85 degrees to sample the intrinsic flattening $c/a$ between 0.13 and 0.73.  The individual grid points sampled are shown in Fig.~3 of the main paper. We note that the observed rotation does not appear to cleanly rotate around the short axis, as the zero-velocity curve appears to twist at 0$\farcs$3. This may suggest that the object is mildly triaxial, but this has minimal impact on our determination of the black hole mass and stellar $M/L$. More significant is the increasing roundness of M60-UCD1 at large radii, which is fitted by our modelling.

The NIFS kinematics are used by the model to constrain the total mass distribution. The confidence contours shown in Figure~3 of the main paper are marginalized over inclination and are based on fits to point symmetrized kinematic data\cite{vandenbosch10}. Error bars are determined for the remaining two degrees of freedom, with $\Delta \chi^2 = 2.30$, 6.18, and 11.83 corresponding to 1$\sigma$, 2$\sigma$ and 3$\sigma$.  The best-fit constant $M/L$ model with no BH has a $\Delta \chi^2=20.0$, and thus is excluded at more than 4$\sigma$.  The reduced $\chi^2$ of the best-fit model to the unsymmetrized data is 0.96 for 280 observables and 3 parameters.  The best-fit inclination is only constrained to be $> 50^\circ$. Such a weak constraint on the viewing angle is usual for dynamical models\cite{vandenbosch09}. There is no dust disk present in this object that can help to constrain the inclination. We note that the maximum $M/L$ ratio expected for an old stellar population with solar metallicity and a canonical IMF is $\sim$4.1 in $V$ band and $\sim$5.1 in $g$ band\cite{mieske13}. This value is somewhat higher than the best-fit $M/L_g \sim 3.6$, but is allowed at $\sim$2$\sigma$ (see Fig.~3 in main paper).  

The ability to detect a black hole with a given set of observations is often quantified by calculating the sphere of influence of the black hole.  While the sphere of influence is normally calculated based on the dispersion of the object, the large BH mass fraction in M60-UCD1 makes the enclosed mass definition of the gravitational sphere of influence $M_{star} (r < r_{infl}) = 2 M_{BH}$ more comparable to previous $r_{infl}$ measurements\cite{merritt13}.  Using this definition, we find $r_{infl} = 0\farcs 27$.  Following the convention presented in a recent review of black hole masses\cite{kormendy13}, we get $r_{infl}/\sigma_{\star} \sim 4$, where $\sigma_{\star}$ is the resolution of our PSF core.  We have many independent measurements of the kinematics within this radius.

The dynamics show that this object has a multi-component structure.  In the phase space there are several components visible (ED~Fig.~3). Roughly 70\% of the stars are on co-rotating orbits, but the remainder are evenly split between counter-rotating and non-rotating, radial, orbits. This indicates that this object was not formed in a single formation event.  The co-rotating orbits dominate at smaller radii with the addition of a non-rotating component at large radii.  This corresponds well to the two component structure fit to the integrated light where we find an inner component with an axial ratio $b/a = 0.75$ and an outer component that is nearly round\cite{strader13}. The anisotropy $\beta \equiv 1-\sigma_{radial}^2/\sigma_{tangential}^2$ is nearly isotropic and does not significantly vary as function of radius. On the other hand $\beta_z \equiv 1-\sigma_{R}^2/\sigma_{z}^2$ gradually decreases outwards from 0 to smaller than -1 and is thus strongly negative.  The anisotropy profiles and orbit types are shown in ED~Fig.~4.  As expected in an (nearly) oblate system it is dominated by short-axis tubes, apart from the region near the black hole where the radial orbits take over.
 
One of the most critical assumptions we make in these best-fit models is the assumption of a constant $M/L$.  This is a well justified simplification in the context of M60-UCD1.  As discussed above in section~3, there is no evidence for any radial color variation, suggesting a stellar population with a constant age.  Furthermore, the formal age estimate from spectral synthesis measurements of integrated optical spectra 14.5$\pm$0.5~Gyr\cite{strader13}, leaving little room for any contribution from young populations with significant $M/L$ differences.  Radial variations in the initial mass function (IMF) that would leave the color unchanged are not excluded {\em a priori} but are highly unlikely, as discussed in the next subsection.

To test our modelling, we also ran Jeans models using the JAM code\cite{cappellari08}.  These models have been shown in the past to give consistent results in comparisons with Schwarzschild models\cite{cappellari10}.  However, we also note that the JAM models enforce a simpler and not necessarily physical orbital structure, including a constant anisotropy aligned with cylindrical coordinates.  Furthermore, the Jeans models do not fit the full line-of-sight velocity distribution as is done with the Schwarzschild models.  JAM models were run over a grid of BH mass, $M/L_g$, anisotropy $\beta$, and inclination $i$.  Fitting only the second moments, $V_{RMS} = \sqrt{V^2+\sigma^2}$, we obtain a best fit stellar dynamical $M/L_g = 2.3 \pm 0.9$ and $M_{BH}=2.4 \pm 0.4 \times 10^7$~M$_\odot$ marginalizing over the other two parameters. The 1$\sigma$ error bars given here were calculated assuming $\Delta \chi^2 < 2.30$ to give comparable error bars to the Schwarzschild models.  
The fit is relatively insensitive to the other two parameters, with $\beta_z$ varying between -1 and 0 and $i$ between 50$^\circ$ and 90$^\circ$.   These models are fully consistent with the Schwarzschild models fits.   A zero BH mass JAM model is strongly excluded in the data with $\Delta \chi^2 > 46$ and a best fit stellar dynamical $M/L_g \sim 6.5$. 

We also use the Jeans modeling to investigate fully isotropic models.  Isotropy is observed in the central parsec of the Milky Way\cite{schodel09}, the only galaxy nucleus where 3D velocities for individual stars have been measured. Furthermore, isotropy has been assumed in the modeling of possible black holes in UCDs\cite{mieske13}.  Interestingly, the $V_{RMS}$ data of M60-UCD1 is also fully consistent with an isotropic model with a $\chi^2 - \chi^2_{min} = 1.1$ (within the 1$\sigma$ contour).  This is consistent with the Schwarzschild models, which find that the system is close to isotropic with $\beta \sim 0.0$.
The best fit isotropic Jeans model has BH mass of $2.2 \pm 0.4 \times 10^7$~M$_\odot$ and $M/L_g = 2.8\pm0.7$. \\

\subsection{Dark Matter and Alternatives to a BH}

We now discuss the possible dark matter content of M60-UCD1 and consider whether an alternative scenario could explain the kinematics of M60-UCD1 without a supermassive BH.    

Dark matter is not expected to make a significant contribution to the kinematics of M60-UCD1. This is due to the extremely high stellar density; dissipationless dark matter cannot achieve anywhere close to the same central densities as baryonic matter. This is shown clearly in previous work matching dark matter halos to galaxies\cite{tollerud11}.  For realistic NFW halos, even a $10^{13}$~M$_\odot$ halo would have only $10^7$~M$_\odot$ within the central 100 pc, while a  $10^{14}$~M$_\odot$ halo would be required to have just $10^6$~M$_\odot$ within M60-UCD1s effective radius; thus even such a massive halo ($\sim$3 orders of magnitude more massive than would be expected for a galaxy with a stellar mass of $\sim$10$^{10}$~M$_\odot$) would contribute only $\sim$1\% to M60-UCD1s mass.  More importantly, a dark matter halo would contribute an extended distribution of mass and fail to produce the central rise in the velocity dispersion.  We note that the one previous resolved measurement of UCD kinematics\cite{frank11} found no evidence for dark matter in a significantly more extended UCD.  Including dark matter in our dynamical models would slightly decrease the stellar $M/L$, which in turn, would further increase the inferred black hole mass in this object.

An alternative to a massive black hole could be a centrally enhanced mass-to-light ratio. To test this scenario we constructed dynamical models without a black hole, but with a radial mass-to-light gradient. We found that the models with an $M/L$ slope of $-0.44\pm0.16$ in log(radius), can yield a good fit, with a $\Delta \chi^2$ difference of $3$ from the best fit constant $M/L + $BH model. This indicates that a model without a black hole and $M/L$ gradient is allowed at $2$-sigma.  However, in addition to providing a slightly worse formal fit to the data, the $M/L$ gradient model is also less physically plausible than the presence of a supermassive BH.  The log(r) dependence means that the center has a very large $M/L$.  This is shown clearly in ED~Fig.~5.  Within our central resolution element (r$\sim$5~pc), the $M/L_g$ in this model is $\sim$12 compared to a value of $\sim$2 at 100~pc.  The model is thus replacing the BH mass with stellar mass near the center in order to match the kinematic data.  

This dramatic $M/L$ gradient cannot be due to stellar age variations given M60-UCD1's uniform color.  The $M/L$ at $r < 5$~pc is a factor of more than 2 above that expected for an old stellar population with a canonical IMF (see above) and thus would require that more than half of the $\sim3.5 \times 10^7$~M$_\odot$ inside that radius be in low mass stars or stellar remnants that produce little light\cite{cappellari12,dabringhausen12}.  We note that dynamical mass segregation is not expected to occur in M60-UCD1, as the half mass relaxation time is $\sim$350~Gyr and remains more than 10~Gyr at smaller radii\cite{binney08,merritt13}.  Thus the only way to explain the $M/L$ gradient would be to have extreme radial variations in the initial mass function.  Assuming a change in the high mass end of the IMF (and thus an increase in stellar remnants), the required upper IMF slope ($dN/dM \propto M^{-\alpha}$) is between 0 and 1.4 at the center, as opposed to the canonical 2.35\cite{dabringhausen09}.  We consider this possibility very unlikely.

\begin{figure}
%\epsscale{0.7}
%\plotone{orbit.eps}
\begin{center}
\includegraphics[width=8.9cm]{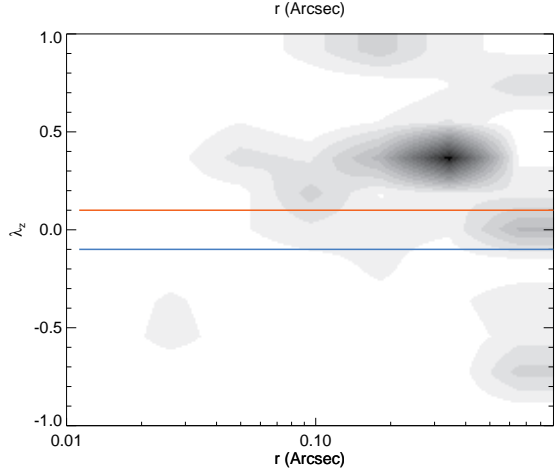}
\end{center}
\caption{Distribution of mass as function of spin $\bar \lambda_z$ and average radius of the orbits as inferred from the  dynamical model. The spin is defined as $\bar \lambda_z = \bar J_z \times (\bar r / \bar \sigma)$, where $\bar J_z$ is the average angular momentum along the $z$-direction, $\bar r$ is the average radius, and $\bar \sigma$ is the average second moment of the orbit. Several distinct components are visisble. While 70\% of the mass is on co-rotating orbits with $\bar \lambda_z \geq 0.1$, there is also a significant amount of mass in components without rotation and counter rotation. }
\label{fig:orbit}
\end{figure}

\begin{figure}
%\epsscale{0.7}
%\plotone{anistropy.eps}
\begin{center}
\includegraphics[width=8.9cm]{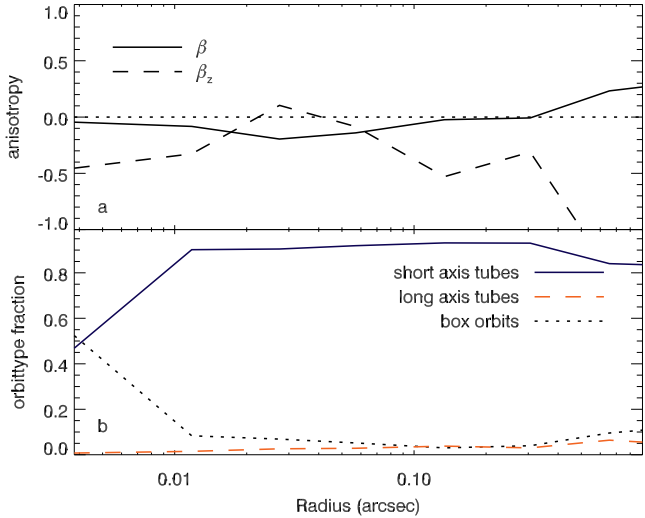}
\end{center}
\caption{Anisotropy and orbit type distribution as function of radius. Top panel shows $\beta$ and $\beta_z$ as solid and dashed line, resp. Anisotropy $\beta \equiv 1-\sigma_{radial}^2/\sigma_{tangential}^2$ indicates the relative size of the velocity ellipsoid in spherical coordinates and it is relative constant over the radii probed by the kinematics. On the other hand  $\beta_z \equiv 1-\sigma_{R}^2/\sigma_{z}^2$ in cylindrical units, gradually declines. Note however that the velocity ellipsoid can not be aligned with the  cylindrical coordinates throughout a stellar system, and thus a physical interpretation of $\beta_z$ is not straightforward. The bottom panel show the relative orbit fraction as function of radius.}
\label{fig:anisotropy}
\end{figure}

\begin{figure}
%\epsscale{1.0}
%\plottwo{M60UCD_mldep_vs_rad.eps}{M60UCD_encmass_vs_rad_bhprof.eps}
\begin{center}
\includegraphics[width=8.9cm]{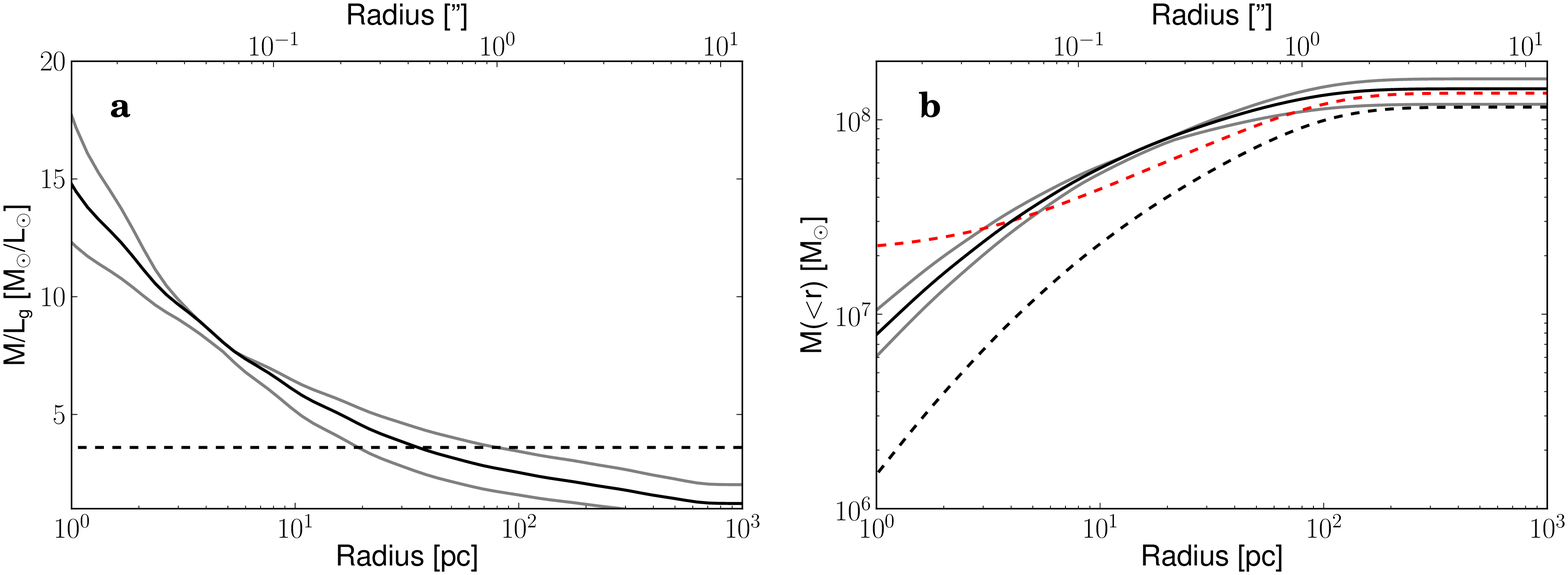}
\end{center}
\caption{{\em Left --} The best fit $M/L$ gradient model (in $g$ band) is shown as the solid line. The maximum $M/L$ for a 13 Gyr population with solar metallicty is $\sim$5.1 in $g$ band, the central $M/L$ in this model is about 3$\times$ this value.  Gray lines show the range $M/L$ gradient fits within 1$\sigma$ of the best fit.  The dashed line shows the best fit constant $M/L$ model for a model including a supermassive BH. {\em Right --} The enclosed mass as a function of radius.  The variable $M/L$ fit with no BH is shown as the solid line with uncertainties in gray.  The black and red dashed line show the enclosed stellar mass and mass including the BH from the constant $M/L$ + BH fit.
}
\label{fig:mass}
\end{figure}

\begin{figure}
\begin{center}
\includegraphics[width=8.9cm]{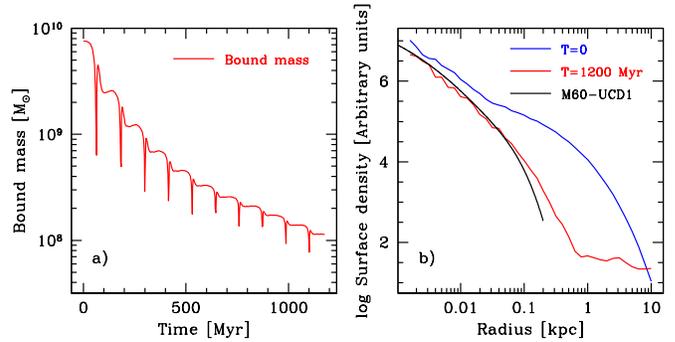}
\end{center}
\caption{{\bf (a)} The evolution of mass bound to the progenitor as it is being stripped to form a UCD.  The progenitor properties were based on estimates from M60-UCD1, while the stripping galaxy is based on the potential of M60. The mass loss is controlled by the distance of pericenter, while the timescale of disruption is set by the apocenter distance.  The pericenter and apocenter in the simulation shown are 1 and 30 kpc respectively.  {\bf (b)} The density profile evolution of tidally stripped galaxy.  The final object has a similar density profile to that of M60-UCD1.  }
\label{fig:strip}
\end{figure}

\vspace{0.3in}

\noindent {\Large {\em Supplementary Information}}

The supplementary information contains further details related to interpretation of our results.  The numbering follows the sections in the methods section.  In section 5 we discuss our calculation of the number density of UCDs with black holes, while in section 6 we show it is feasible for M60 to disrupt a $\sim10^{10}$~M$_\odot$ galaxy to form an object like M60-UCD1.  We also discuss alternative formation scenarios.

\section{Number Density of UCDs with Black Holes in Fornax}

\hspace{1in}

In the main text we state that a comparable number of black holes may be found in UCDs as are found in current galaxy nuclei.  We base this estimate on number counts of UCDs and galaxies in the Fornax cluster, where the catalog of UCDs is most complete. We note that the UCD catalog in e.g. Virgo is very incomplete, since most UCDs have been found within the boundaries of archival $HST$ data\cite{brodie11} and with even extremely massive UCDs still being found\cite{strader13}.

The Fornax UCDs were identified in spectroscopic surveys\cite{drinkwater00,gregg09} and are complete to an $M_V \sim -10.7$, which corresponds to a mass of $\sim 5 \times 10^6$~M$_\odot$.  Assuming masses from SSP models, there are a total of 23 UCDs above $10^7$~M$_\odot$ and a total of 170 UCDs between $2 \times 10^6$ and $10^7$~M$_\odot$\cite{hilker11}.  The sample is likely complete within 320~kpc of the NGC~1399 for the higher mass bin, but likely underestimates the total number of lower mass UCDs by $\sim$30\%\cite{mieske12}.  

We now try to estimate how many of these UCDs may host massive BHs.  As discussed in the main manuscript, the stellar dynamical $M/L$ derived for M60-UCD1 ($M/L_g = 3.6 \pm 1.0$, giving $M/L_V = 2.8 \pm 0.8$) is lower than the dynamical $M/L_V$ estimates of 18 of the 19 UCDs with mass estimates above 10$^7$~M$_\odot$ and about half the UCDs between 3$\times$10$^6$~M$_\odot$ and 10$^7$~M$_\odot$\cite{mieske13}.  In the high mass bin, 15 of 19 are more than 1$\sigma$ above this value, while in the low mass bin, this fraction is 25\%. This is despite the fact that the stellar $M/L_V$ is expected to increase with increasing metallicity, and with its solar metallicity\cite{strader13}, M60-UCD1 is among the most metal-rich UCDs. Furthermore, M60-UCD1 is among the densest UCDs and thus might be expected to have a high stellar $M/L$ under scenarios where IMF variability is tied to the star formation rate density\cite{dabringhausen12}.  Therefore, we assume that the high dynamical $M/L$ estimates seen in UCDs are due to massive BHs\cite{mieske13}.  If we take the fractions of objects more than 1$\sigma$ above the M60-UCD1's $M/L$, then we have $\sim$18 BHs in the high mass sample, and an additional $\sim$56 in the low mass sample.  Thus, we predict $\sim$74 BHs in the Fornax cluster.

We can then compare this number to the likely number of BHs that reside at the centers of galaxies.  Although the occupation fraction of BHs in low mass galaxies is poorly constrained, X-ray observations of AGN in early type galaxies show almost no detections in galaxies with a stellar mass below $\sim$3$\times 10^{9}$~M$_\odot$\cite{gallo10}.  Modeling of this data shows that the occupation fraction is still high at this mass but likely starts to drop at lower masses\cite{greene12,miller14}.  We therefore try to estimate the number of galaxies above $3 \times 10^9$~M$_\odot$ in the Fornax cluster to get a rough estimate of the number of nuclear black holes in Fornax.  Using the catalog of likely Fornax cluster candidates\cite{ferguson89b}, which should be complete at the magnitudes of interest, we derive stellar mass estimates using available luminosities and colors from HYPERLEDA\cite{paturel03} and derive an $M/L$ estimates based on the colors\cite{bell03}.  We find a total of 45 galaxies in Fornax with stellar mass above $3 \times 10^9$~M$_\odot$ (of which 37 are early type galaxies) that are likely to host BHs.  We note that just 5 Fornax cluster galaxies have reliable dynamical BH mass determinations\cite{mcconnell13a}.  

If we consider just the BHs in higher mass UCDs, they represent a $\sim$40\% increase over the number of galaxy BHs in Fornax, while inclusion of the lower mass UCDs could more than double the number of BHs.  While high mass $>10^7$~M$_\odot$ UCDs are found primarily in galaxy clusters, lower mass UCDs with BHs may be present in group environments where a majority of galaxies live.  If high $M/L$ ratio UCDs do indeed have BHs, they will make a substantial contribution to the total number of BHs in the local universe.

\section{Plausibility of Tidal Stripping by M60 and Alternative Formation Scenarios}

\hspace{1in}

In order to test whether M60-UCD1 could have formed from tidal stripping,
we simulated the orbital evolution of the UCD inside M60 through a
series of direct N-body simulations using NBODY6\cite{aarseth99}. We
used N=$2 \times 10^5$ particles for the UCD plus its progenitor
galaxy, while M60 was modelled as a constant background potential.
This potential included a $4\times10^9$~M$_\odot$ central supermassive BH plus
the sum of two NFW components for stars and dark matter closely
resembling the previously published mass profile\cite{shen10}.  

The initial progenitor galaxy was assumed to be the M60-UCD1 nuclear
component (S\'ersic profile with $r_{eff}=14$ pc, $n=$3.3 and
mass$=7.4\times10^7$~M$_\odot$) plus a galaxy described by a single
S\'ersic component with $r_{eff}=1.15$~kpc, $n=2$.0 and total
mass$=10^{10}$~M$_\odot$.  These values were based on elliptical
galaxies with known nuclear star cluster components similar to the
inner component of M60-UCD1 and are also broadly consistent with the
BH bulge mass relation as discussed in the main paper.  Some specific
examples galaxies with similar nuclear star clusters are NGC~4379 \&
NGC~4387 in Virgo\cite{cote06} and NGC~1389 \& IC~2006 in
Fornax\cite{turner12}.  These galaxies have $M_B = -18$ to $-19$ and
stellar mass estimates from $1-3 \times 10^{10}$~M$_\odot$.  These
specific early-type galaxies and other similar galaxies in this
luminosity range have S\'ersic indices of
2-3\cite{kormendy09,ferrarese06b} and dispersions of
50-80~km/s\cite{koleva11}.

Stripping of the galaxy to form an object similar to M60-UCD1 requires
a pericenter of $\lesssim$1.5~kpc; with an apocenter of 30~kpc or more
the dynamical friction inspiral timescale of the resulting remnant
becomes larger than the Hubble Time.  Thus a very radial orbit is required if M60-UCD1 was stripped a long time ago as its old stellar population might suggest.  Results of the total mass evolution and surface brightness profile evolution in one representative simulation is shown in ED~Fig.~6. 

The formation of hypercompact stellar systems has also been suggested from gravitational recoil of merging black holes\cite{merritt09b}.  These predicted systems can have sizes and luminosities similar to UCDs, but few systems as bright as M60-UCD1 are expected, and such systems should have integrated dispersions $\gtrsim$6 times higher than are observed for M60-UCD1.  Alternatively, the detection of a high mass fraction black hole in NGC~1277\cite{vandenbosch12} suggests direct formation of very massive black holes might be possible.  However, the mass estimate of the BH in NGC~1277 may be somewhat overestimated\cite{emsellem13}, and there is limited theoretical understanding of how such high mass fraction black holes could be produced\cite{bonoli14,shields13}. Thus, we consider tidal stripping of a massive progenitor galaxy as the most likely formation scenario for M60-UCD1.

\begin{table}
\begin{center}
\small
\begin{tabular}{ccc}
\hline \hline
$\rm L_\odot \rm pc^{-2} $  &  log $\sigma '$ & $q'$ \\ 
(mag)    &  (arcsec) &  \\ 
 \hline    
      337.276 &   -1.88135 &   0.999 \\
      651.189 &   -1.33255 &   0.999 \\
      928.698 &  -0.918423 &   0.999 \\
      909.443 &  -0.593469 &   0.999 \\
      554.490 &  -0.337421 &   0.999 \\
      196.581 &  -0.131899 &   0.999 \\
      36.1820 &  0.0420669 &   0.999 \\
      2.26539 &   0.207857 &   0.999 \\
      374202. &   -3.50825 &   0.749 \\
      334899. &   -3.08561 &   0.749 \\
      274935. &   -2.69367 &   0.749 \\
      204099. &   -2.32475 &   0.749 \\
      130962. &   -1.98010 &   0.749 \\
      74045.5 &   -1.66058 &   0.749 \\
      36353.7 &   -1.35852 &   0.749 \\
      14894.5 &   -1.07469 &   0.749 \\
      5126.86 &  -0.808290 &   0.749 \\
      1457.43 &  -0.558353 &   0.749 \\
      342.689 &  -0.323720 &   0.749 \\
      65.1802 &  -0.102248 &   0.749 \\
      10.0077 &   0.107907 &   0.749 \\
      1.17965 &   0.316168 &   0.749 \\
    0.0766351 &   0.553877 &   0.749 \\
\hline
\end{tabular}
\end{center}
\caption{Luminosity model of M60-UCD1 composed of 23 Gaussians based on the F475W-band \emph{HST}/ACS image. The position angle of the galaxy is -49.45$^\circ$. Col.~(1): apparent surface brightness, with a correction for galactic foreground extinction (0.087 mag)\cite{schlafly11}, and assuming a solar luminosity in $g$ band of 5.12 Mag.  Col.~(2): size along the major axis. Col.~(3): flattening}
\end{table}

\end{document}